# THE FIRST SYNCHROTRON INFRARED BEAMLINES AT THE ADVANCED LIGHT SOURCE: SPECTROMICROSCOPY AND FAST TIMING

MICHAEL C. MARTIN* AND WAYNE R. MCKINNEY
Advanced Light Source, Lawrence Berkeley National Laboratory
Berkeley, CA 94720 USA

Two recently commissioned infrared beamlines on the 1.4 bending magnet port at the Advanced Light Source, LBNL, are described. Using a synchrotron as an IR source provides three primary advantages: increased brightness, very fast light pulses, and enhanced far-IR flux. The considerable brightness advantage manifests itself most beneficially when performing spectroscopy on a microscopic length scale. Beamline (BL) 1.4.3 is a dedicated FTIR spectromicroscopy beamline, where a diffraction-limited spot size using the synchrotron source is utilized. BL 1.4.2 consists of a vacuum FTIR bench with a wide spectral range and step-scan capability. This BL makes use of the pulsed nature of the synchrotron light as well as the far-IR flux. Fast timing is demonstrated by observing the pulses from the electron bunch storage pattern at the ALS. Results from several experiments from both IR beamlines will be presented as an overview of the IR research currently being done at the ALS.

## INTRODUCTION

Over the past two years a new infrared beamline and three experimental endstations have been commissioned at the Advanced Light Source, Lawrence Berkeley National Laboratory. They have been designed[1] to use the specific advantages gained when a synchrotron light source is used for infrared and UV[2] spectroscopy. A schematic diagram of the IR beamline facility is shown in Figure 1. The bending magnet synchrotron light is reflected from a plane mirror, M1, followed by an ellipsoidal mirror, M2, and finally a third flat mirror, M3, all in ultra high vacuum (UHV). The UHV ends with a wedged diamond window just after M3 and near the image made by M2. The light then proceeds in rough vacuum inside the "switchyard" where it is collimated by two cylinders and sent towards one of the three experimental endstations. Details can be found in Ref. 1.

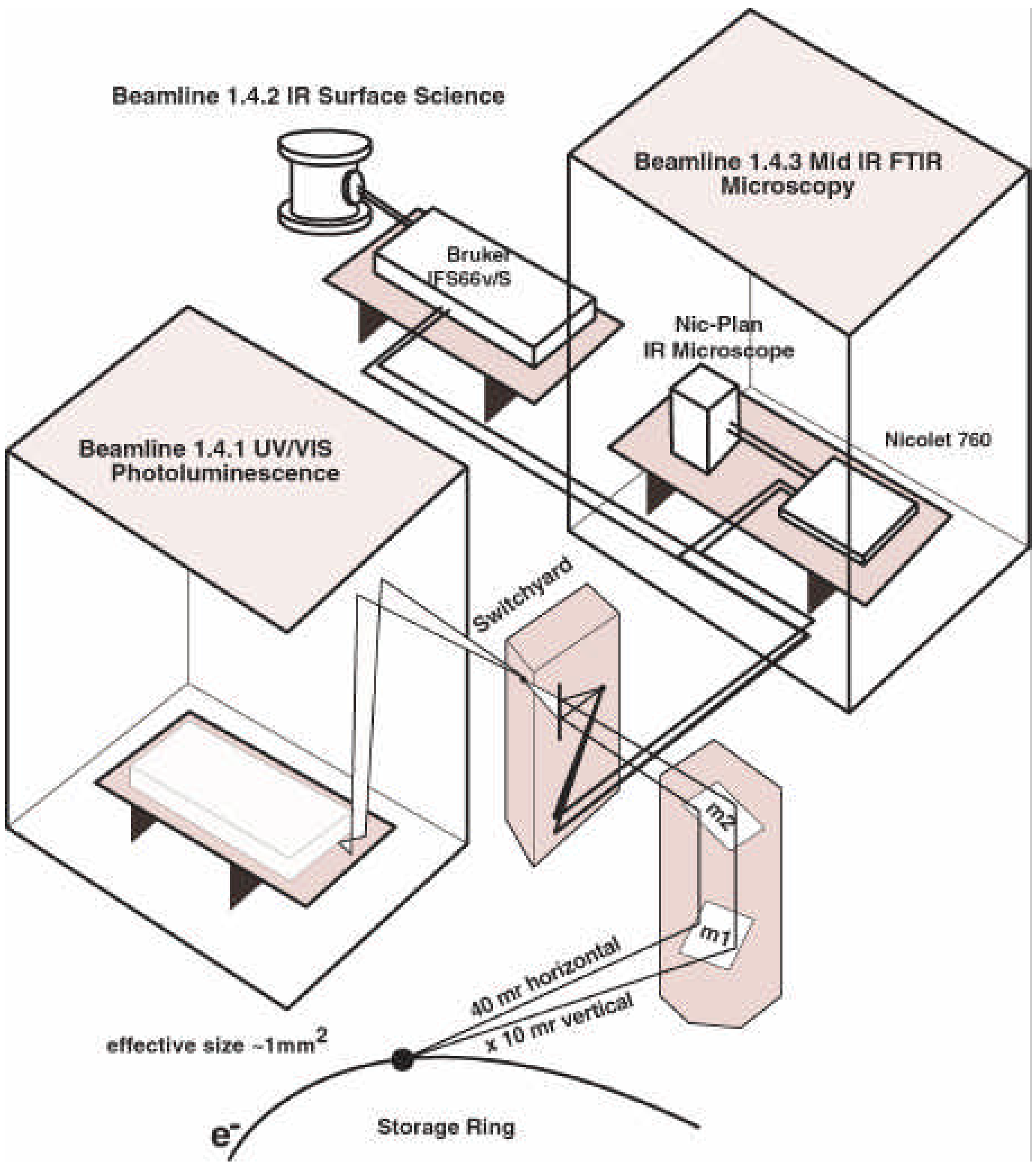

FIGURE 1 Schematic diagram of the ALS infrared beamline complex.

BEAMLINE 1.4.3: FTIR SPECTROMICROSCOPY

The collimated synchrotron light beam exits the switchyard (see Figure 1) and then enters the beamline 1.4.3 hutch through a hole in the wall all in rough vacuum. The beam exits the vacuum via a KBr crystal window and then enters into the side of a Nicolet 760 FTIR bench. Once inside the bench, the light goes through the FTIR Michelson interferometer, then is sent on to the sample which can be either in the main bench's sample compartment or in a Spectra-Tech Nic-Plan IR microscope, and finally to an IR detector.

One of the primary reasons for doing infrared spectroscopy at a synchrotron light source is the large enhancement in brightness (flux per unit area). This brightness advantage manifests itself most beneficially when focussing the light to a very small spot size[3]. We have been able to achieve essentially diffraction-limited spot sizes in the mid-IR using the Nic-Plan microscope with the ALS source. Figure 2 demonstrates the tight focus achieved using the synchrotron source, a 32x objective, and measuring the transmission through a 5-micron pinhole on the sample stage. No other apertures were used. The pinhole was moved via the computer controlled x-y stage to map out the extent of the focused beam spot. The cross sectional Gaussian widths of the spot were 6.2 by 10.10 microns. Since the mid-IR wavelengths used are centered in the 3 to 10 micron range, and the objective lenses have a numerical aperture of approximately 1, we would expect the diffraction limited spot size to be approximately the wavelength of light, i.e. around 10 microns or better. This has been achieved on BL1.4.3. When the same measurement is made using the internal Globar$^{TM}$ IR source, the spot widths are 80 by 100 microns! Thus while the synchrotron source does not provide a huge gain in mid-IR flux compared to the Globar$^{TM}$, we are able to get *all* of the light from the synchrotron through a 10-micron aperture. We routinely measure a signal enhancement with a 10-micron sample of a factor of 200.

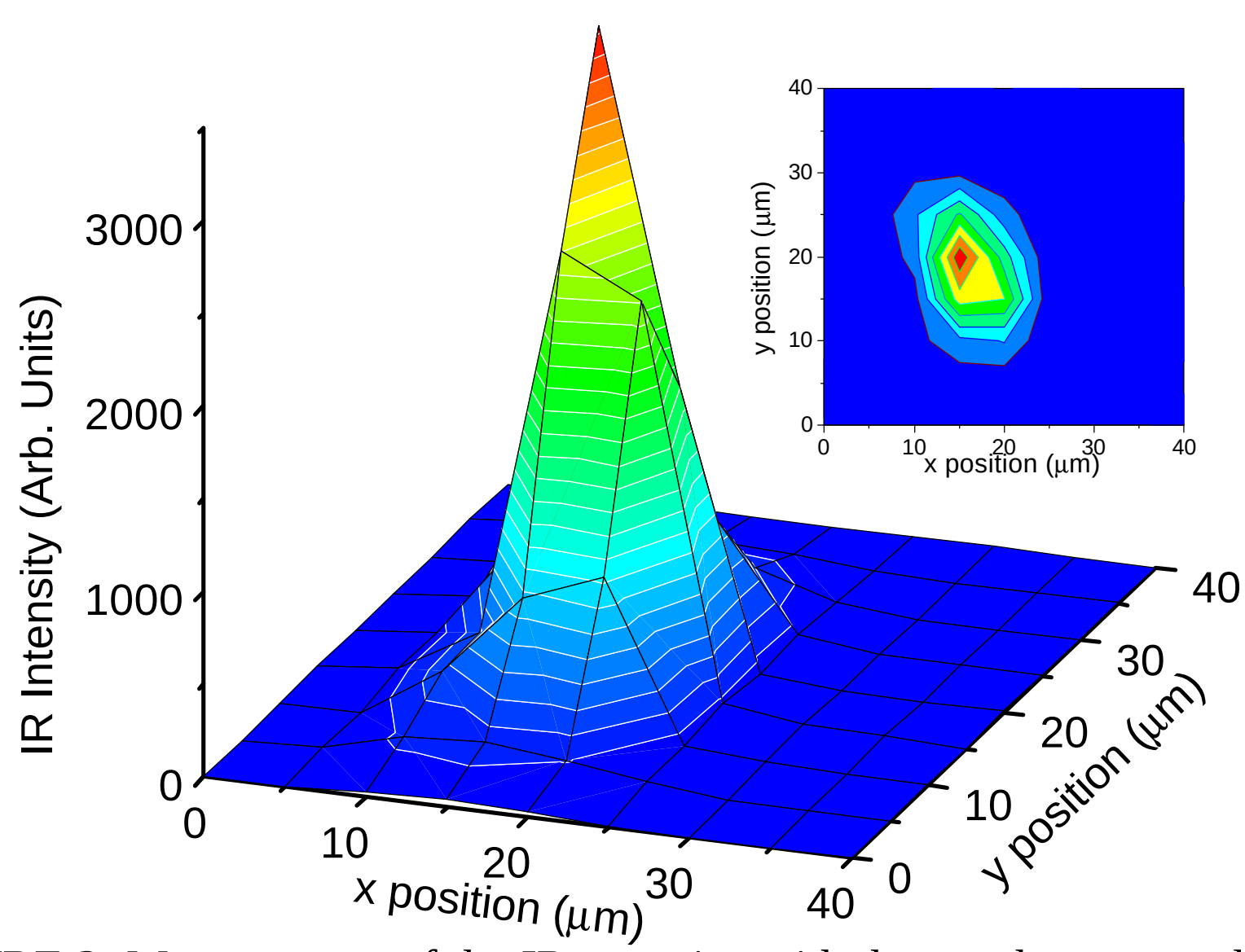

FIGURE 2 Measurement of the IR spot size with the synchrotron as the light source demonstrating the essentially diffraction-limited spot size.

Other equipment available for users of BL1.4.3 include an MMR micro miniature regfrigerator[4] that allows sample measurements at temperatures between 70K to 740K, an autofocus capability for the IR microscope, and a grazing-incidence objective for greater surface sensitivity.

We currently have many beamline users spanning a number of scientific disciplines. Figure 3 shows one example of what can be done with the small IR spot size available at BL1.4.3. The work of Holman *et al*.[5] has shown that small (10-20 micron) colonies of specific bacteria can reduce highly toxic Chromium in the VI oxidation state to the much less soluble, and therefore much less toxic trivalent chromium. When toluene is present as a co-contaminant the Cr(VI) reduction is observed to proceed quicker, presumably because the toluene is used as a carbon source. Figure 3 shows three reflectance maps all taken from the same spectral data set of a bacteria colony living on rock. The colony of bacteria is identified by the presence of the well known amide II peak of protein, and the absence of both toluene and Cr(VI) peaks at the same location as the colony strikingly shows the potential of the synchrotron-based FTIR technique. Obtaining a comparable signal to noise with the internal source would make the local depletion of Cr(VI) and toluene impossible to observe, and would eliminate the specific identification of the chromium reduction with the colony of bacteria.

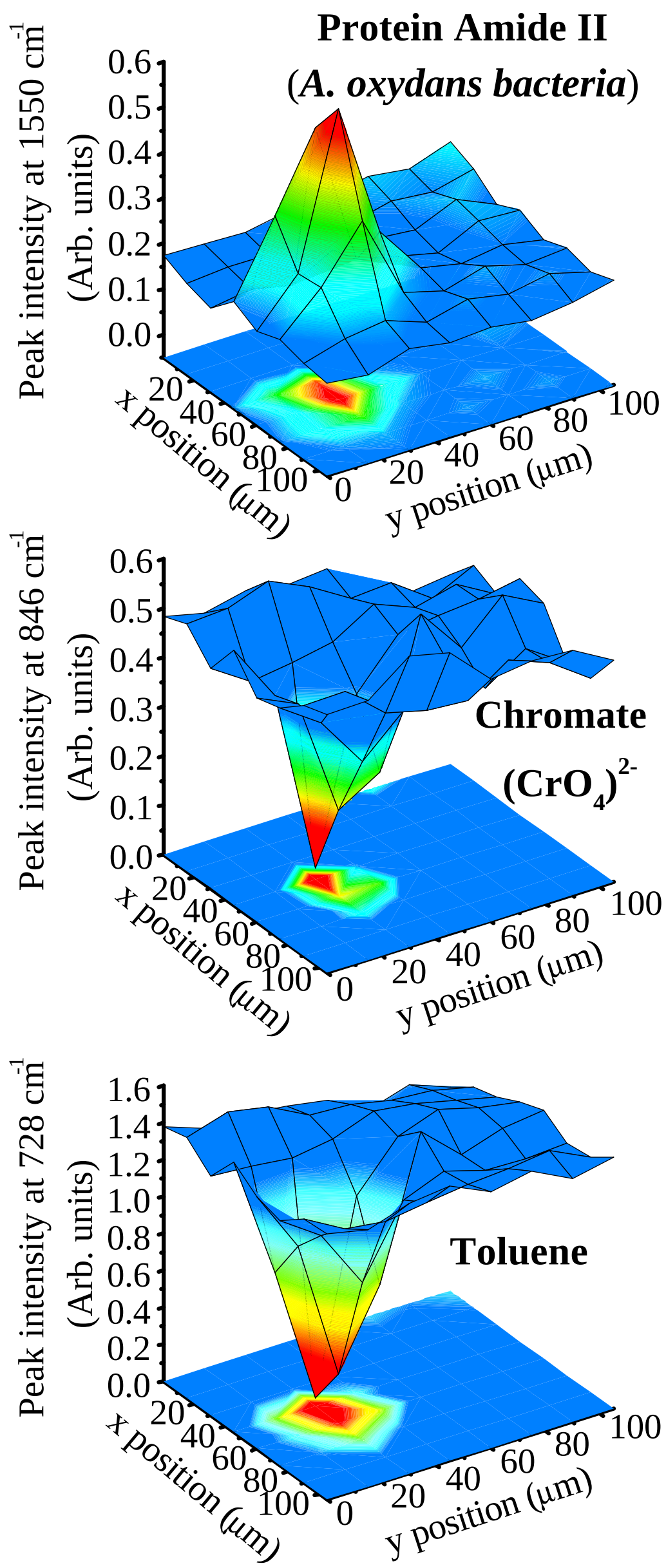

FIGURE 3 Anthrobacter oxydans bacteria, isolated from a contaminated DOE site in Idaho, attach themselves to magnetite mineral surfaces. We locate the bacteria colony via their spectral signature (top). We observe a depletion of chromate (middle) and toluene (bottom) by the bacteria after five days of exposure.

## BEAMLINE 1.4.2: VACUUM FTIR SPECTROSCOPY

The collimated synchrotron light can bypass the Nicolet spectrometer described above and continue to beamline 1.4.2 (see Figure 1). This experimental endstation consists of a Bruker *IFS* 66v/S vacuum FTIR spectrometer that is capable of measuring from 20 to 25,000 $cm^{-1}$, and can be run in step-scan mode. The modulated light can pass in vacuum to the regular sample compartment for either transmission of reflection measurements, or it can exit the spectrometer and refocused onto a surface in a UHV chamber for grazing incidence spectroscopy of sub- to several monolayer surfaces. The main sample chamber can also be fitted with a Janis liquid helium cryostat for measurements of samples at temperatures of 1.4 to 475K.

Measurements in the UHV chamber gain from the high-brightness of the synchrotron compared to a standard IR source because of the excellent focusing capabilities when using the synchrotron IR beam. Additionally because this spectrometer can measure into the far-infrared region, the enhancement of far-IR flux from a synchrotron can be put to use. Finally, this spectrometer can also make use of the pulsed nature of the synchrotron beam.

Standard rapid scan Fourier Transform Infrared Spectroscopy (FTIR) acquires scans on the time scale of one second, so the very fast pulses of a synchrotron source are not noticed. However, using the step-scan capabilities and fast electronics of the *IFS* 66v/S FTIR bench on BL1.4.2, we can measure IR spectra with a time resolution as fast as 5 nsec. This means that processes that occur on the nanosecond all the way up to hour time scales can be monitored with time-resolved spectroscopy using this beamline.

To demonstrate the fast-timing capabilities of this beamline, we synchronized the detection electronics with the ALS ring timing structure (one trigger per complete revolution of the electrons). The ALS typically operates in multibunch mode with 288 bunches spaced 2 nanoseconds apart, followed by an 80 nsec gap. The individual pulse width is ~44 picoseconds (FWHM). This electron-filling pattern is schematically drawn in Figure 4.

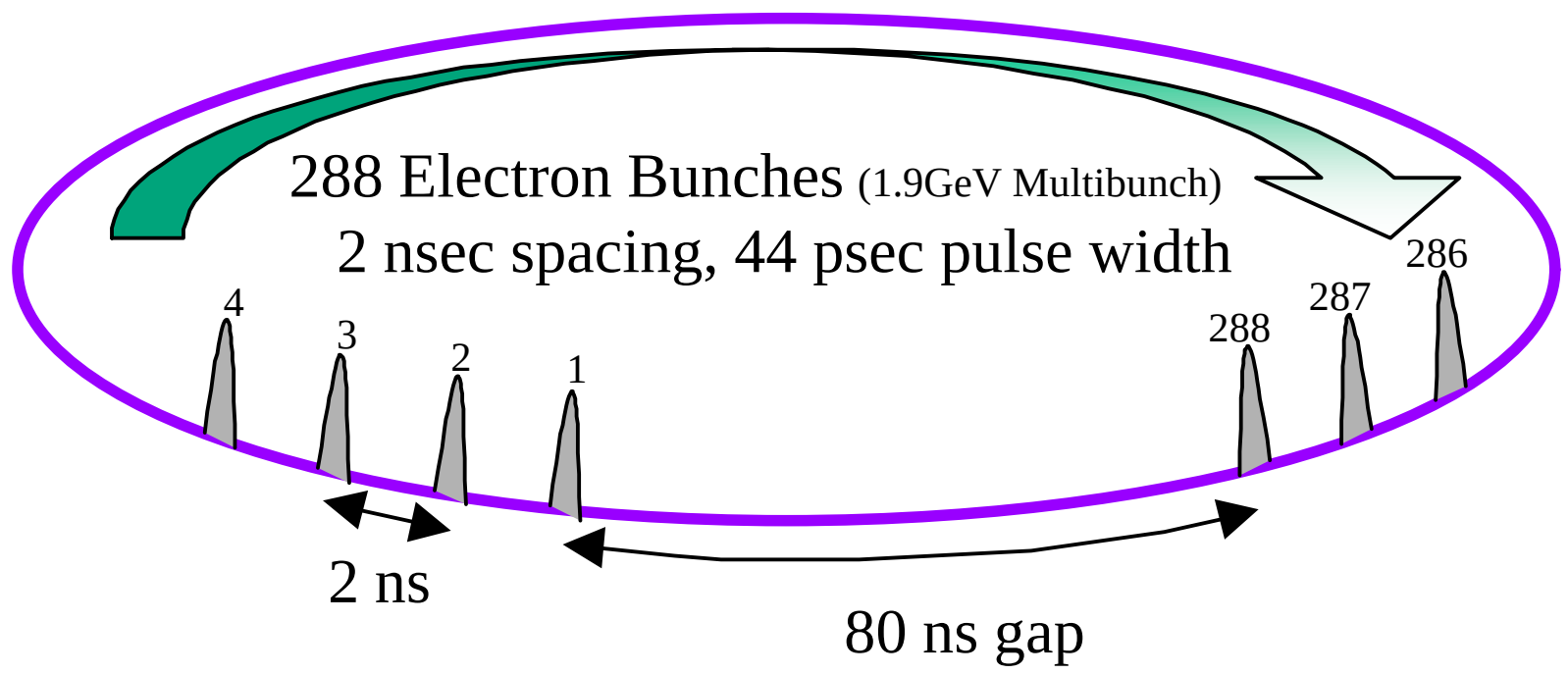

FIGURE 4 The ALS generates photon pulses at intervals of 2nsec, except for a single long gap of 80ns. We use this time structure to demonstrate the fast timing capabilities of BL1.4.2.

IR spectra were obtained at 5nsec time slices for a total of 650 nsec (a complete revolution of each electron bunch). We plot the measured intensity as a function of time integrated over a region in the visible wavelengths in Figure 5. This figure clearly demonstrates that we can observe the 80-nsec gap in the synchrotron light pulses. The IR detector and digitization electronics are not yet fast enough to observe the 2-nsec pulse spacing, but forthcoming enhancements should allow sub-nanosecond timing enabling the use of individual synchrotron pulses to serve as the probe in very fast time-resolved experiments.

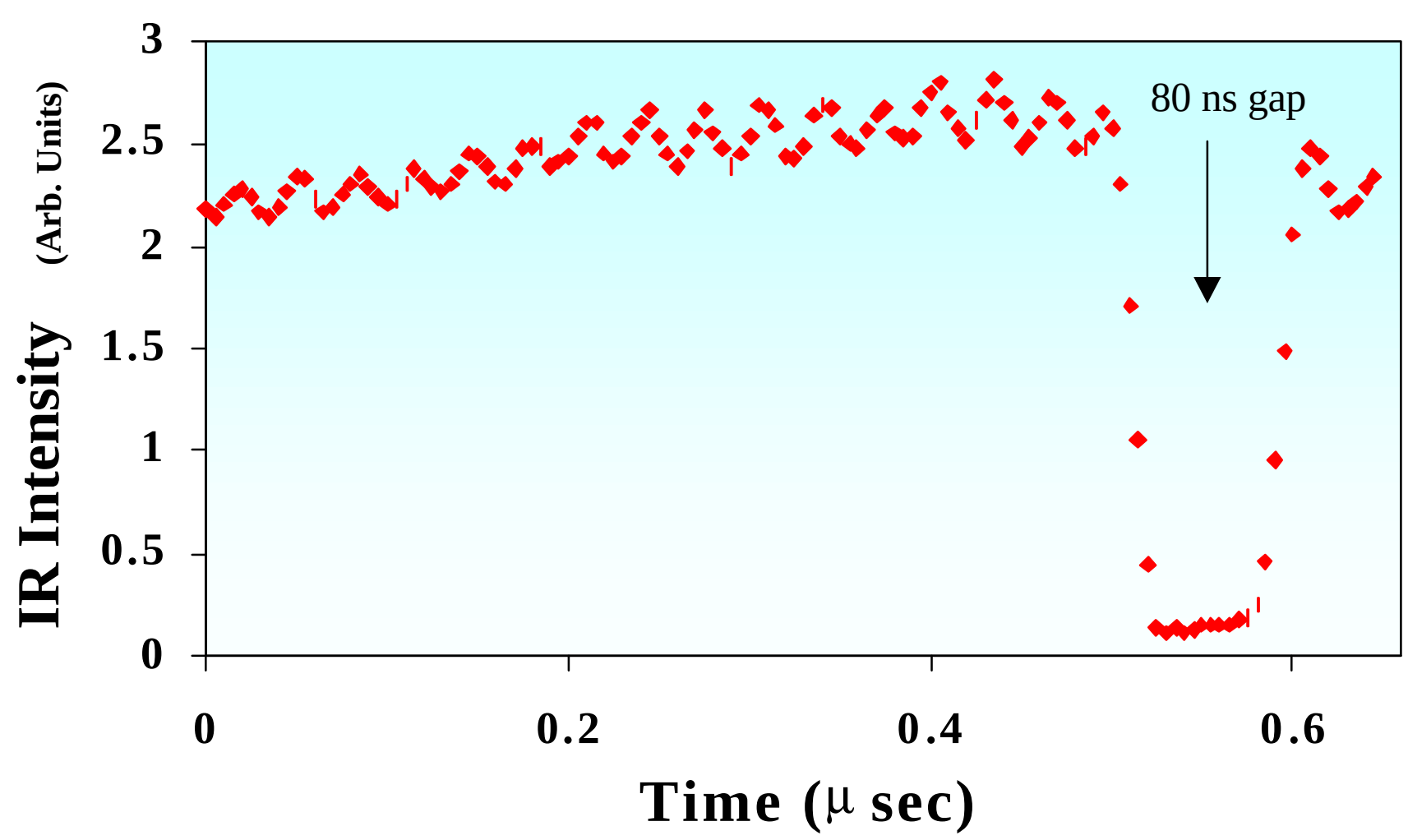

FIGURE 5. Time-resolved measurement of the IR intensity showing the 80-nsec gap in the ALS synchrotron filling pattern. The time resolution of the current electronics does not yet fully separate the 2-nsec spacing between pulses.

The vertical aperture limit of 10 mrad (full angle) at the storage ring aperture should be the limiting factor in how far the long wavelength range will extend. To push the far-IR capabilities of the ALS we are exploring several possibilities. The edge radiation from the sector zero injection straight may be a much better way to extract far IR from the ALS given the vertical opening angle constraints. See, for example, the paper and cited references in Bosch.[6] Another possibility is making a small storage ring with very short electron bunch lengths.[7] This will allow coherent addition of wavelengths longer than the electron bunch length and therefore a significant increase in the far-IR flux.

## BEAMLINE 1.4.1: UV/VIS PHOTOLUMINESCENCE

The left side of Figure 1 shows the UV/VIS photoluminescence area. M1 and M2 are coated with bare aluminum. This coating with its natural oxide layer helps absorb extreme UV which would normally pass gold IR mirrors,[1] but preserves reflectivity to about 6 volts.[2] For use of 1.4.1, M3 is lowered allowing the light to pass into an additional low vacuum box welded onto the switchyard where the light is collimated with an off-axis paraboloid. Standard UV/VIS components, for example, quartz lenses, monochromators and spectrographs are used on the optical table. Flux approaching calculated values (including lens absorption, etc) are obtained.

## NOISE REDUCTION

Reducing noise in the FTIR spectra of the microscope has been the major challenge in the commissioning of the beamlines. Initial mid-IR RMS noise measurements taken by dividing two sets of 64 spectra at a resolution of 4 $cm^{-1}$ from reflection off a gold coated microscope slide in October 1997 showed the intolerable value 0.5%. Even though a several day study of vibration in the area was done prior to the construction of the beamline, two 20-hp pumps, which circulated water for the RF cavities in the storage ring, were not operating at the time of the testing. The pumps drove the shield walls and floor of the ALS at their fundamental rotation (58 Hz) frequency and at the fourth multiple of that frequency. In addition the low conductivity water (LCW) supplied by the ALS was plumbed near the lines for the RF chiller water, and picked up the same frequencies. Additional higher frequency noise was found on the electron beam itself due to a noisy master oscillator for the ALS. The systematic reduction of vibration and electron beam noise problems is characterized in Byrd *et al.*[8] We have succeeded in reducing the typical noise a factor of ten to twenty to the range of <0.05% RMS which we believe is typical of the day to day noise level at synchrotron-based IR facilities.

In order to continue improving the noise level we are implementing an active feedback system based on Hamamatsu position sensitive detectors, Physik Instrumente two-axis tip/tilt piezo mirror mounts and custom feedback circuitry. The details of this feedback system will be described in a later publication.

## CONCLUSIONS

A tremendous signal to noise advantage is provided by the brightness of the synchrotron source providing for a real, significant advantage for spectromicroscopy studies in the mid-IR wavelength range at high spatial resolutions. Absorption spectra from single human cells and bacteria are now routinely possible. The ALS IR beamlines have been expanded for uses in the far-IR for the investigation of High-$T_c$ and related oxide systems, and step scan and timing abilities have been added. It is expected that one or two IR surface science chambers will be routinely connected to the Bruker bench where in grazing incidence the synchrotron IR brightness advantages will be again put to use. The UV end of the spectrum from the 1.4 front end is now used for photoluminescence and other related measurements by a third end station.

ACKNOWLEDGEMENTS

This work was supported by the Director, Office of Energy Research, Office of Basic Energy Sciences, Materials Sciences Division of the U.S. Department of Energy, under Contract No. DE-AC03-76SF00098.

REFERENCES

1. W. R. McKinney, C. J. Hirschmugl, H.A. Padmore, T. Lauritzen, N. Andresen, G. Andronaco, R. Patton, and M. Fong, SPIE Proceedings, **3153**, pp. 59-76, (1997). (LBNL-40848, UC-410, LSBL-414)
2. J.W. Ager III, K.M. Yu, W. Walukiewicz, E.E. Haller, Michael C. Martin, W.R. McKinney, and W. Yang, J. Appl. Phys., **85**, 8505, (1999). (LBNL-42638)
3. Jean Louis Bantignies, Larry Carr, Paul Dumas, Lisa Miller, and Gwyn P. Williams, Synchrotron Radiation News, **11**, (1998).
4. MMR Technologies, Inc., 1400 North Shoreline Blvd., # A5, Mountain View, CA 94043, http://www.mmr.com/
5. Hoi-Ying N. Holman, Dale L. Perry, Michael C. Martin, Geraldine M. Lamble, Wayne R. McKinney, and Jennie C. Hunter-Cevera, Geomicrobiology J., **16**(4), 1999, *in press*.
6. R.A. Bosch, Nucl. Inst. Meth. in Phys. Res., A **431** pp. 320-333, (1999).
7. J.B. Murphy and S. Krinsky, Nucl. Inst. Meth. in Phys. Res., A **346** pp. 571-577 (1994).
8. J.M. Byrd, M. Chin, M.C. Martin, W.R. McKinney, and R. Miller, Proceedings of the SPIE Annual Meeting, Denver, June 1999.

---

* E-mail: MCMartin@lbl.gov, IR Beamline Web Page: http://infrared.als.lbl.gov/